\documentstyle[preprint, eqsecnum, tighten,aps,pstricks,]{revtex}
\begin{document}
\title{Quenched, Minisuperspace, Bosonic $p$-brane Propagator}
\author{Stefano Ansoldi\footnote{E-mail
address: ansoldi@trieste.infn.it}}
\address{Dipartimento di Fisica Teorica\break
Universit\`a di Trieste\\
INFN, Sezione di Trieste}
\author{Antonio Aurilia\footnote{E-mail
address: aaurilia@csupomona.edu}}
\address{Department of Physics\\
	California State Polytechnic University\break Pomona, CA 91768 }
\author{Carlos Castro\footnote{E-mail
address: castro@ctsps.cau.edu    }}
\address{Center for Theoretical Studies of Physical Systems\break
Clark Atlanta University\break Atlanta, GA 30314 }
\author{Euro Spallucci\footnote{E-mail
address: spallucci@trieste.infn.it}}
\address{Dipartimento di Fisica Teorica\break
Universit\`a di Trieste\\
INFN, Sezione di Trieste}
\maketitle
\begin{abstract}
We borrow the {\it minisuperspace approximation} from Quantum Cosmology and
the
{\it quenching approximation} from $QCD$ in order to derive a new form of
the bosonic $p$--brane
propagator. In this new approximation we obtain an {\it exact}
description of both the {\it collective mode} deformation of the brane
and the center of mass dynamics in the target spacetime. The collective
mode dynamics is a generalization of string dynamics in terms of area
variables. The final result is that the evolution of a $p$--brane in the
quenched--minisuperspace approximation
is formally equivalent to the effective motion of a
particle in a spacetime where {\it points as well as
hypersurfaces} are considered on the same footing as fundamental
geometrical objects. This geometric equivalence leads us to define a new
{\it tension--shell condition} that is a direct extension of the
Klein--Gordon condition for material particles to the case of a physical
$p$--brane.

\end{abstract}

\newpage

	\section{Introduction}
	The purpose of this paper is to introduce a new approximation
	scheme to the quantum dynamics of extended objects. Our approach
differs from the more conventional ones, such as the normal modes expansion
or higher
	dimensional gravity, in that it is inspired by two different
quantization schemes:
	one is the {\it minisuperspace} approach to Quantum Cosmology (QC),
the other is the
	{\it quenching} approximation to $QCD$. Even though these
schemes apply
	to different field theories, they have a common rationale, that
is, the idea of quantizing only a finite number of degrees of freedom while
	freezing, or ``quenching,'' all the others.
	In QC this idea amounts, in practice, to quantizing a single scale
factor (thereby
	selecting a class of cosmological models, for instance, the
	Friedman--Robertson--Walker
	spacetime) while neglecting the quantum fluctuations of the full
metric. The effect is to turn the exact, but intractable, Wheeler--DeWitt
	functional equation \cite{wdw} into an ordinary quantum
	mechanical wave equation \cite{padma}.
	As a matter of fact, the various forms of the ``wave function of the
	universe'' that attempt to describe the quantum birth of the cosmos are
	obtained through this kind of approximation \cite{hh},
	or modern refinements of it \cite{gianni}. On the
	other hand, in $QCD$, the dynamics of quarks and gluons cannot be
solved
	perturbatively outside the small
	coupling constant domain. The strong coupling regime is usually
	dealt with by studying the theory on a lattice. However, even in
that case,
	the computation of the fermionic determinant by Montecarlo simulation
	is actually impossible. Thus, in the ``quenched approximation,'' the
	quark determinant is set equal to unity, which amounts to
neglecting the effect of
	virtual quark loops. In other words, this extreme approximation in
terms of heavy--quarks with a vanishing number of flavors assumes
	that gauge fields affect quarks while quarks have no dynamical
	effect on gauge fields \cite{quench}.\\
\\
	Let us compare the above two situations with the basic problem that one
faces when dealing with
	the dynamics of a {\it relativistic extended object,} or $p$-brane,
	for short. Ideally, one would like to account for all {\it local}
deformations of the object configuration, i.e., those deformations that may
	take place at a generic point on the $p$--brane. However, any
attempt to provide a local
	description of this shape--shifting process leads to a
	functional differential equation similar to the Wheeler--DeWitt
	equation in QC. The conventional way of handling that functional
equation, or the
	equivalent infinite set of ordinary differential equations
	\cite{mramond},
	is through perturbation theory. There, the idea is to quantize
	the small oscillations about a classical configuration and
	assign to them the role of ``particle states'' \cite{hosocar}.
	Alternative to this approach is the quantization of a brane of
preassigned
	geometry. This (minisuperspace) approach, pioneered in
Ref.(\cite{ctuck}), was used in Refs.(\cite{minib},
	\cite{minib2}) in order to estimate the nucleation rate of a
	spherical membrane. Presently, the minisuperspace approximation is
introduced for the purpose of providing an {\it exact} algorithm for
computing two specific components of the general dynamics of a $p$--brane:
one is the brane {\it collective} mode of oscillation in terms of global
volume variations, the other is the evolution of the brane center of mass.\\
\\
In broad terms, this paper is divided into two parts: Section II deals with
classical dynamics; Section III deals with quantum dynamics in terms of the
path--integral, or ``sum over histories.'' \\
Since the action for a classical $p$--brane is not unique, we start our
discussion by providing the necessary background with the intent of
justifying our choice of action. We then take the first step in our
approximation scheme in order to separate the center of mass motion from
the bulk and boundary dynamics. In subsections IIC and IID we derive an
effective action for the bulk and boundary evolution, while in subsection
IIE we discuss the meaning of the ``quenching approximation'' at the
classical level.\\
An approximation scheme for a dynamical problem is truly meaningful and
useful only when the full theory is precisely defined, so that the
technical and logical steps leading to the approximate theory are clearly
identified. Thus, in Section III we first tackle the problem of computing
the general quantum amplitude for a $p$--brane to evolve from an initial
configuration to a final one. The full quantum propagator is obtained as a
sum over all possible histories of the world--manifold of the relativistic
extended object. In subsection IIIA, we show in detail what that ``sum over
histories'' really means, both mathematically and physically, in order to
explain why the bulk quantum dynamics cannot be solved exactly. What {\it
can} be calculated, namely, the boundary and center of mass propagator, is
discussed in subsections IIIB and IIIC. The final expression for the
quantum propagator and the generalized
``tension--shell'' condition in the quenched--minisuperspace approximation
is given in
subsection IIID. Finally, subsection IIIE checks the self--consistency of
the result against
some special cases of physical interest.

	\section{Classical Dynamics}

	\subsection{Background}

The action of a classical $p$--brane is not unique.
	The first (mem)brane action, dates back to 1962 and was introduced by
	Dirac in an attempt to resolve the electron--muon puzzle
\cite{dng}. The Dirac action
	was reconsidered in Ref.(\cite{ctuck}) and quantized following the
pioneering
	path traced by Nambu--Goto in the lower dimensional string
	case. The Dirac--Nambu--Goto action represents the world volume
	of the membrane trajectory in spacetime. Thus, it can be generalized
	to higher dimensional $p$--branes as follows

\begin{equation}
S_{\mathrm{DNG}}\left[\, Y\, \right] =
-m_{p+1}\int_{\Sigma _{p+1}} d^{ p+1} \sigma\, \sqrt{ -\gamma}
\quad , \quad
\gamma \equiv \det \left(\, \partial_m \, Y^\mu \, \partial_n \, Y_\mu\,\right)
\ ,
\label{ng}
\end{equation}
where $m_{p+1}$ represents the ``$p$--tension'' ( we denote with ``$p$'' the
spatial dimensionality of the brane ) and the coordinates $\sigma^m$,
$m=0, 1, \dots{}, p$, span the $(p+1)$--dimensional world--manifold
$\Sigma$ in parameter space. On the other hand, the embedding  functions $Y
^\mu(\sigma)$, $\mu=0, 1, \dots{}, D-1$,
represent the
brane coordinates in the target spacetime.\\
An alternative description that preserves the reparametrization
invariance of the world--manifold is achieved by introducing an auxiliary
metric $g_{ m n}(\sigma)$ in parameter space together with a ``cosmological
constant'' on the world--manifold\cite{ht},
\cite{netra}

\begin{equation}
S_{\mathrm{HTP}}\left[\, Y\ , g\, \right]= -{m_{p+1}\over 2}\,\int_\Sigma
d^{ p+1}\sigma\, \sqrt{ -g}\,
\left[\, g^{ m n}\partial_m\, Y^\mu\, \partial_n\, Y_\mu -(p-1)\,\right]
\ ,
\label{spolya}
\end{equation}
where $g\equiv \det\, g_{ m n}$. The two actions {  (\ref{ng})} and
{   (\ref{spolya})}
are classically equivalent in the sense that the ``field equations''
${\delta S/\delta g^{ m
n}(\sigma)}=0$ require the auxiliary world metric to match the
induced metric, i.e., $g_{ m n}=\gamma_{ m n}=\partial_m\,
Y^{\mu}\, \partial_n\, Y_{\mu}$. The two actions are also complementary:
$S_{\mathrm{DNG}}$ provides an ``extrinsic'' geometrical description in terms
of the embedding functions $Y^\mu(\, \sigma\, )$ and the induced metric
$\gamma_{mn}$, while $S_{\mathrm{HTP}}$ assigns an ``intrinsic'' geometry to
the world manifold $\Sigma$ in terms of the metric $g_{ m n}$ and the
``cosmological constant'' $m_{p+1}$, with the $Y^\mu(\, \sigma\, )$
functions interpreted
as a ``multiplet of scalar fields'' that propagate on a curved
$(p+1)$--dimensional manifold.\\
Note that in both functionals {  (\ref{ng})}
and {  (\ref{spolya})}, the brane tension $m_{p+1}$ is a {\it
pre-assigned}
parameter. More recently, new action functionals have been proposed that
bridge the gap between relativistic extended objects and gauge
fields\cite{gauge},
\cite{gauge2}, \cite{gauge3}.
The brane tension itself, or world--manifold cosmological constant, has been
lifted from an {\it a priori} assigned parameter to a dynamically generated
quantity that may attain both positive and vanishing
values. Either a Kaluza--Klein type mechanism \cite{blt} or a modified
integration measure have been proposed as
dynamical processes for producing tension at the classical \cite{eduard}
and semi-classical level \cite{sec}.\\
For our present purposes, the form {  (\ref{spolya})} of the
$p$--brane action is the more appropriate
	starting point.
	There are essentially two reasons for this choice:
    \begin{enumerate}
	\item Unlike the Nambu--Goto--Dirac action, or the Schild action
\cite{schild}, Eq.{  (\ref{spolya})} is quadratic in the variables
$\partial_m X^\mu$. As we shall see in the following subsections, this
property, together with the choice of an appropriate coordinate system
	on $\Sigma_{p+1}$, facilitates the factorization of the center
	of mass motion from the deformations of the brane.

	\item Equation {  (\ref{spolya})} can be interpreted as a
	{\it scalar field theory in curved spacetime}. From this point
	of view, the minisuperspace quantization approach is
	equivalent to a quantum field theory in a fixed background
	geometry, at least as far as the auxiliary metric is concerned.
  	    \end{enumerate}

	\subsection{Center of Mass Dynamics}

	The dynamics of an extended body can be formulated in general as
	the composition of the center of mass motion and the motion
	relative to the center of mass. \\
	A $p$-brane is by definition a spatially extended object. Thus we
expect
	to be able to separate the motion of its center of mass from the
	{\it shape--shifting} about the center of mass. However, given the
point like nature of the center of mass, its spacetime coordinates depend
on one parameter only, say, the proper time $\tau$. Thus, the factorization
of the
        center of mass motion automatically breaks the general
        covariance of the action in parameter space since it breaks the
symmetry between the temporal parameter $\tau$ and the spatial coordinates
$s ^i$.
	We can turn ``needs into  virtue'' by choosing a coordinate mesh on
$\Sigma_{p+1}$ that reflects the breakdown of general covariance in
parameter space.
Indeed, we can choose the model manifold $\Sigma_{p+1}$ of the form
	\begin{equation}
	\Sigma_{p+1}=I\otimes  \Sigma_p\quad , \quad \partial I=\{ P_0 ,
	P\}\quad ,\quad \partial  \Sigma_p=\emptyset,
	\end{equation}
	where $I$ is an {\it open } interval of the real axis,
	which has two points, say $P _{0}$ and $P$, as its boundary and
$\Sigma_p$
	is a finite volume, $p$--dimensional manifold, without boundary.
	Thus, $\partial \Sigma_{p+1}= P_0\otimes\Sigma_p \cup
	P\otimes\Sigma_p$ and the spacetime image of $\partial
	\Sigma_{p+1}$ under the embedding $Y$ represents the initial
	and final brane configuration in target spacetime.

In terms of coordinates, the above factorization of $\Sigma_{p+1}$ amounts
to defining $\tau$ as
	the center of mass proper time and the $ s_i$'s as spatial
coordinates of $\Sigma_p$. Accordingly, the invariant line
	element reads:

	\begin{equation}
	dl^2= \overline g_{\, mn} \, d\sigma^m\,  d\sigma^n=
	 -e^2(\, \tau\, )\,\, d\tau^2 +h_{ij}(\, \vec s\, )\, \,
	 ds^i\, \, ds^j
	\label{ds1}
	\end{equation}

where $\tau$ plays the role of ``cosmological time'', that is, all
	clocks on $\Sigma_p$ are synchronized with the center of
	mass clock.\\
	Now, we are in a position to introduce
	the {\it center of mass coordinates} $x^\mu(\tau)$ and the
	{\it relative coordinates} $Y^\mu(\, \tau , s^i\, )$:
	 \begin{eqnarray}
          X^\mu(\, \tau\ , \vec s\, )
		 &\equiv&
		 x^\mu(\tau) +{1\over\sqrt{m_{p+1}}}\, Y^\mu(\, \tau\ , \vec s
		 \,)\ ,
		 \label{xcm}\\
         x^\mu(\tau)
		 &\equiv&
		 { 1\over V_p} \int_{{ {S}}_p} d^p s\,
         \sqrt{h(\,\vec s\, )}\,  X^\mu(\, \tau\ , \vec s\, )\\
		 V_p
		 &\equiv&
		 \int_{{ {S}}_p} d^p s\, \sqrt{h(\, \vec s\, )}\ , \qquad
		 h(\, \vec s\, )\equiv det\left(\, h_{ij}\,\right)\ .
	\end{eqnarray}

	Using the above definitions in the action {  (\ref{spolya})} and
	replacing  $g_{mn}$ with $\overline g_{\, mn}$ as indicated in
	Eq.(\ref{ds1}), we find

	 \begin{eqnarray}
	 S &=& - {1\over 2}\int_\Sigma d^{p+1}\sigma\,
	 \sqrt{\, \overline g\, }\, \left[ \, m_{p+1}\, {\overline g}^{\, 00}\,
	 \dot x^\mu( \tau )\, \dot
	 x_\mu( \tau ) + \left(\, {\overline g}^{\, m n}\,
	\partial_m\,  Y^\mu\,
	\partial_n\,  Y_\mu - m_{p+1}\, (p-1)\, \right)\, \right]\ ,\qquad
	p\ge 1
	\nonumber\\
	&=& -{1 \over 2}\, m_{p+1}\, V_p \,\int_0^T \!\!\!\!\!d\tau \left[\,
	-{\dot x^\mu(\tau)\, \dot x_\mu(\tau)\over e(\tau) }
	+ e(\tau)\, \right]
	-{1\over 2}\, \int_{\Sigma _{p+1}} \!\!\!\!\!\!\!d^{p+1}\sigma\,
	 \sqrt {\, \overline g\, }\,
	\left[\, \overline{g}^{\, m n}\, \partial_m\, Y^\mu\,
	\partial_n \, Y_\mu -m_{p+1}\, p\, \right]\ .
	\end{eqnarray}
	The first term describes the free motion of the bulk center of
	mass. The absence of a mixed term, one that would couple the center
	of mass to the bulk oscillation modes, is due to
	the vanishing of the metric component $\overline g_{\, 0i}$ in the
        adopted coordinate
	system {  (\ref{ds1})}. The last term represents the usual
        bulk modes free
	action for a covariant ``scalar field theory'' in parameter space.\\
	Finally, if we define
	the brane {\it volume mass,} $M_0  \equiv V_p \, m_{p+1}$,
	representing the brane inertia under volume variation, then, from
the above expression, we can read off the center of mass action  and the
corresponding Lagrangian

	\begin{equation}
	 S_{cm}=-{M_0\over 2}  \int _0^T d\tau\,
	\left[\,
	 -{\dot x^\mu\, \dot x_\mu\over e(\tau)  }
	 +   e(\tau) \right]\equiv
	 \int _0^T d\tau\,  L_{cm}\left(\,  \dot x^\mu\ ; e(\tau)\, \right)\ ,
	 \end{equation}
	where the {\it einbein} $e(\tau)$ ensures $\tau$
	reparametrization invariance along the center of mass world--line.\\
        Summarizing, the final result of this subsection is that, in the
	adopted coordinate frame where the center of mass motion is
        separated from the bulk and boundary dynamics, we can write the total
        action as the sum of two terms

	\begin{equation}
	S=S_{cm}  -{1\over 2}\, \int_{\Sigma_{p+1}} d^{ p+1}\sigma\,
         \sqrt{\, \overline g\, }\,
        \left[\, \overline{g}^{\,  m n}\, \partial_m \, Y^\mu\,
        \partial_n\,  Y_\mu -m_{p+1}\, p\, \right]\ .
        \label{ssplit}
	\end{equation}

	We emphasize that, in order to derive the expression {
       (\ref{ssplit})}, it was necessary to break the full invariance under
	general coordinate transformations of the initial theory, preserving
	only the more restricted symmetry under independent time and spatial
	coordinate reparametrizations.

\subsection{Induced Bulk and Boundary Actions}

	In this subsection we wish to discuss those features of the
	brane classical dynamics which are instrumental for the subsequent
	evaluation of the quantum path--integral.\\
	In agreement with the restricted reparametrization invariance of
       the action { (\ref{ssplit})}, as discussed in the previous subsection,
	we first set up a ``canonical formulation'' which preserves that
        same symmetry
	through all computational steps. This means that all the
	world indices $m\ , n\ , \dots$ are raised, lowered and contracted
	by means of the center of mass metric $\overline{g}_{\, mn}$.
	From the brane action {  (\ref{ssplit})} we extract the brane
	{\it relative momentum} $P^m{}_\mu$ and the corresponding
	Hamiltonian $H$ :

	\begin{eqnarray}
	P^m{}_\mu & \equiv &
	{\partial L\over \partial  \partial_m\, Y^\mu(\, \sigma\, )}= -
	\overline{g}^{\, mn}\, \partial_n\, Y_\mu(\, \sigma\, )\,
	\label{mom}
	\nonumber \\
	H & \equiv & P^{ m}{}_\mu\, \partial_m\, Y^\mu -L
	=  -{1\over 2}\, \left[ \, \overline{g}_{ \, m n}\, P^{
	m}{}_\mu\, P^{n}{}^\mu + m_{p+1} \, p\, \right]\ .
	\label{covshild}
	\end{eqnarray}

	Thus, we can write the action in the following canonical form

	\begin{eqnarray}
	S & = & \int_{\Sigma_{p+1}} d^{ p+1}\sigma\, \sqrt{\, \overline g\,} \,
	\left(\, P^{ m}{}_\mu \, \partial_m \, Y^{\mu} - H\, \right)
	\nonumber\\
	& = & \int_{\Sigma_{p+1}} d^{ p+1}\sigma\, \sqrt{\, \overline g\, }\,
	\left[\, P^{ m}{}_\mu\, \partial_m\, Y^{\mu}+ {1\over 2}\,
	\left(\, \overline{g}_{\, ab}\, P^{ a}{}_\mu \, P^{ b}{}^\mu   +
	p\, m_{p+1}\, \right)  \,  \right].
	\label{sfase}
	\end{eqnarray}

	 The first term in Eq.(\ref{sfase}) can be rewritten as follows

	\begin{equation}
	\int_{\Sigma_{p+1}} d^{ p+1}\sigma\, \sqrt{\, \overline g\, }\,
	P^{ m}{}_\mu\, \partial_m\, Y^{\mu}=
	\int_{\Sigma_{p+1}} d^{ p+1}\sigma\, \left[\,
	\partial_m\, \left( \, \sqrt{\overline g} \,  P^{ m}{}_\mu\,
	Y^{\mu}\, \right)-Y^{\mu}\, \partial_m\, \left( \,\sqrt{\,\overline
g\,}
	\, P^{ m}{}_\mu\, \right)\, \right]\ .
	\label{bulbound}
	\end{equation}
	According to Eq. {  (\ref{bulbound})} we can write the total
action
	as the {\it sum of a boundary term plus a bulk term}
	\begin{eqnarray}
	S & = & S_B[\, \partial {\Sigma_{p+1}} \, ] + S_J[\, {\Sigma_{p+1}}
	\, ]\nonumber\\
	& = & \int_{\Sigma_{p}} d^{p} s \, \sqrt{h}\, N _{n}\,  p^n{}_\mu\,
	 y^\mu -\int_{\Sigma_{p+1}}
	d^{p+1}\sigma \,\sqrt{\,\overline g\, }\, Y^\mu(\, \sigma\, )\,
	\nabla_m\, P^{ m}{}_\mu - \int_{\Sigma_{p+1}} d^{p+1}\sigma\,
	\sqrt{\, \overline g\, }\, H\ ,
	\end{eqnarray}
	where $ p^n{}_\mu$ and $y^\mu$ are the momentum and coordinate
	of the boundary, $ d^{p} s \sqrt{h} N _{n}$ represents the oriented
surface
	element of the boundary, and $\nabla_m$ stands for the covariant
derivative with
	respect to the metric $\overline{g}_{\, mn}$.
	The distinctive feature of this rearrangement is that {\it the
	bulk coordinates $ Y^\mu(\sigma)$  enter the action as Lagrange
	multipliers enforcing the classical equation of motion}:

	\begin{equation}
	{\delta S\over\delta \, Y^\mu(\, \sigma\, ) }=0
	\quad \Longrightarrow \quad \partial_m\, \left(\,
	\sqrt{\, \overline g\, }\, P^{m}{}_\mu\, \right)=0\ .
	\label{transv}
	\end{equation}

	The general solution of Eq.{  (\ref{transv})} may be expressed
as follows

	\begin{eqnarray}
	 P^{ m}{}_\mu
	 &=&
	 \overline{g}^{ \, mn}\,\partial_n\, \phi_\mu +
	 {1\over p!}\, \overline{\epsilon}^{ \, mm_2\dots m_{p+1}}
	 \left(\,
	 P^{0)}{}_{\mu\mu_2\dots \mu_{p+1}}\, \partial_{m_2}\, Y^{ \mu_2}\dots
	 \partial_{m_{p+1}}\,
	 Y^{ \mu_{p+1}} +
	 \partial_{[\, m_2}\, A_{\mu m_3 \dots m_{p+1}\, ] }\,\right)
	 \nonumber\\
	 &=&
	 \overline{g}^{ \, mn}\, \partial_n \, \phi_\mu +
	 {1\over p!}\, \overline{\epsilon}^{ \, mm_2\dots m_{p+1}}\,
	\left( \, P^{0)}{}_{\mu m_2\dots m_{p+1}}+
	F(\, A\, )_{\mu m_2\dots m_{p+1}}\, \right)
    	\end{eqnarray}

     	with

    	\begin{eqnarray}
	\Box_{\overline g} \,  \phi^\mu(\, \sigma\, )&=&0\,\\
	\partial_m \, P^{0)}{}_{\mu\mu_2\dots \mu_{p+1}} &=&0\ .
	\label{solp}
	\end{eqnarray}

	 Here, the components $\phi_\mu$ represent local {\it harmonic
modes} on the bulk, and
	$\overline{\epsilon}^{\, mm_2\dots m_{p+1}}\equiv (\overline
        g)^{-1/2}\, \delta^{[\, m  m_2\dots m_{p+1}\, ]}$ stands for the
totally
	antisymmetric tensor. Moreover, the constant antisymmetric tensor
	$ P^{0)}{}_{\mu\mu_2\dots \mu_{p+1}}$ represents the  volume momentum
	{\it zero--mode,} or {\it collective--mode}, that describes the
{\it global}
	volume variation of the brane.

	In order to be able to treat $\phi$, $P^{0)}$ and
	$A$ as independent oscillation modes, we demand that the following
        orthogonality relations are satisfied:

	\begin{eqnarray}
	{\mathcal{C}} _{\mathrm{I}} &\equiv &
	\int_{\Sigma_{p+1}} d^{p+1}\sigma\, \sqrt{\, \overline g\, }\,
	\overline{\epsilon}^{\, m  m_2\dots m_{p+1}}\, \partial_m \,\phi^\mu\,
	\partial_{m_2}\, Y^{\mu_2}\dots \partial_{m_{p+1}}\, Y^{\mu_{p+1}} =0,
	\label{ort1} \\
	{\mathcal{C}} _{\mathrm{II}} &\equiv &
	\int_{\Sigma_{p+1}} d^{p+1}\sigma \,\sqrt{\,\overline g\,}
	\overline{\epsilon}^{mm_2\dots m_{p+1}} \,\partial_m\, \phi^\mu\,
	\partial_{[\, m_2}\, P ^{0)} _{\mu m_2 \dots m_{p+1}\, ]} =0\ ,
	\label{ort2} \\
	{\mathcal{C}} _{\mathrm{III}} &\equiv &
	\int_{\Sigma_{p+1}} d^{p+1}\sigma\, \sqrt{\,\overline g\,}\,
	\overline{\epsilon}^{\, mm_2\dots m_{p+1}}\, \partial_m\, \phi^\mu\,
	\partial_{[\, m_2}\, A_{m_3 \dots m_{p+1}\, ]} =0\ .\label{ort3}
	\end{eqnarray}
	The three orthogonality constraints on the bulk determine the
	field behavior on the boundary through Stokes' theorem. In particular
    	${\mathcal{C}} _{\mathrm{I}}$ gives

	\begin{equation}
	{\mathcal{C}} _{\mathrm{I}}=\int_{\Sigma _p}\, \phi^\mu \,
	dy^{\mu_2}\wedge \dots\wedge
	dy^{\mu_{p+1}}=0\quad\Longrightarrow\quad \phi^\mu(\,\vec s\, )=0,
	\label{b1}
	\end{equation}

	which is a Dirichlet boundary condition, whereas in
	${\mathcal{C}} _{\mathrm{III}}$ two integrations remain:

	\begin{eqnarray}
	{\mathcal{C}} _{\mathrm{III}} &=& \int_{\Sigma _{p}} d^{p}s \,
	 F_\mu{}^{m_2\dots m_{p+1}}\, y^\mu\partial_{m_2}\, y^{\mu_2}\dots
	 \partial_{m_{p+1}}\, y^{\mu_{p+1}}
	\nonumber\\
	& & -\int_{\Sigma_{p+1}} d^{p+1}\sigma\, \sqrt{\, \overline g\, }
	\left(\,
	\partial_{m_2}\, F_\mu{}^{m_2 m_3\dots m_{p+1}}\, Y^\mu\,
	\partial_{ m_3}\, Y^{\mu_3}\dots \partial_{ m_{p+1}}\, Y^{\mu_{p+1}}
	\right)\ .
	\label{b2}
	\end{eqnarray}

	Since the two integrations are carried over the boundary and
	the bulk respectively, the orthogonality condition can
	be satisfied only if each integral is identically vanishing,
	\begin{eqnarray}
	 {\mathcal{C}} _{\mathrm{III}} =0 \quad &\Longrightarrow&
	 A^\mu{}_{m_3\dots m_{p+1}}(\, \vec s\, )=\partial_{[\, m_3 }\,
	 \Lambda^\mu\, {}_{m_4\dots m_{p+1}\, ] }
	 \nonumber \\
	 & &\mathrm{and} \quad \partial_{m_2}\, F_\mu{}^{m_2 m_3\dots
m_{p+1}}=0
	 \ .
	\end{eqnarray}
	Thus, $A$ must solve free Maxwell--type equations on the bulk
	and reduce to a pure gauge configuration on the boundary.
	Note that under these
	conditions for $\phi$ and $A$, ${\mathcal{C}} _{\mathrm{II}}$ is
	satisfied as well.\\
	In summary, the classical solution for the brane momentum reduces
	the original field content of the model to:
	\begin{itemize}
	\item a multiplet $\phi^\mu$ of world, harmonic, scalar
	fields (target spacetime vector) which satisfy Dirichlet boundary
	conditions;
	\item a multiplet $A^\mu{}_{m_3\dots m_{p+1} }$ of world,
	Kalb--Ramond fields (target spacetime vector) which reduce to a pure
	gauge configuration on the  boundary;
	\item a world--manifold (cosmological) constant
	$P^{0)}{}_{\mu\mu_2\dots \mu_{p+1}}$
	(target spacetime constant tensor) corresponding to
	a constant energy background along the brane world--manifold.
	\end{itemize}

\subsection{Effective Bulk and Boundary Actions}

	By inserting Eq.{  (\ref{solp})} into the action {
(\ref{sfase})}, and
        taking into account the conditions {  (\ref{ort1})}, {
(\ref{ort2})},
        {  (\ref{ort3})},  {  (\ref{b1})}, {  (\ref{b2})},
         we can write an {\it effective} classical action for the three
        types of oscillation modes,

\begin{eqnarray}
	 S^{eff} &\equiv& S_B + S_J \label{jeff}\\
	 S_B &=& {1\over (p + 1) !}\, P^{0)}{}_{\mu\mu_2\dots\mu_{p+1}}\,
	\int_{\partial\Sigma}d\sigma^{\mu\mu_2\dots\mu_{p+1}}(\, \vec s\, )+
	 \int_{\Sigma_{p}} d^p s \, y^\mu \,  N^m(\, \vec s\, )\,
	  \partial_m\, \phi_\mu\\
	  &=& {1\over (p + 1) !}\, P^{0)}{}_{\mu\mu_2\dots\mu_{p+1}}\,
	\sigma^{\mu\mu_2\dots\mu_{p+1}}+
	 \int_{\Sigma_{p}} d^p s \, y^\mu \,  N^m(\, \vec s\, )\,
	  \partial_m\, \phi_\mu\\
	S_J&=&-{1\over 2 m_{p+1}}\, \int_{\Sigma_{p+1}} d^{p+1}\sigma\,
	\sqrt{\, \overline g\, }
	\left(\,
	 \overline{g}^{\, mn}\, \partial_m \, \phi^\mu\, \partial_n \, \phi_\mu
	 -{1\over p!}\,  F^\mu{}_{m_2\dots m_{p+1}}\, F_\mu{}^{m_2\dots
m_{p+1}}
	\, \right)
	\nonumber\\
	&&\qquad +{1\over 2 m_{p+1}}\,
	\left[\, {1\over (p+1)!} P^{0)}{}_{\mu m_2\dots m_{p+1}}\,
	P^{0)}{}^{\mu m_2\dots m_{p+1}}
	 - m_{p+1}^2\, p\, \right]\, \Omega_{p+1}\ ,
	 \label{sclass}
	  \end{eqnarray}

where $N^m(\, \vec s\, )$ represents the normal to the boundary and

	  \begin{equation}
	 \sigma^{\mu\mu_2\dots\mu_{p+1}} =\int_{\partial\Sigma}\,
	y^\mu\, dy^{\mu_2}\wedge\dots\wedge dy^{\mu_{p+1}}\equiv
	\int_{\partial\Sigma}d\sigma^{\mu\mu_2\dots\mu_{p+1}}(\,\vec s\, )\ ,
	\qquad p\ge 1
	\label{sigmadef}
	\end{equation}

stands for the volume tensor of the brane
	 in target spacetime, while $d\sigma^{\mu\mu_2\dots\mu_{p+1}}(\, \vec
	 s\, )$ represents the {\it oriented volume element} attached to
the original $p$--brane at the contact point $x^\mu=y^\mu(\,\vec s\,)$.
Finally, by definition, we set

	   \begin{equation}
	  \Omega_{p+1}\,\equiv \int_{\Sigma_{p+1}} d^{ p+1}\sigma \,
	  \sqrt{\overline g}  =
	  \int_0^T d\tau\, e(\tau)\, \int_{\Sigma_{p}} d^p\sigma\,
	\sqrt h \equiv V_p\,  \int_0^T d\tau\,  e(\tau)\ .
	   \label{ipervol}
	   \end{equation}

	    Expression {  (\ref{sigmadef})} allows us to establish
	  a relation between {\it functional } derivatives in
$p$--\textit{loop space},

	  \begin{equation}
	  {\delta\over \delta y^\mu( \,\vec s \, )}=
	  y^{\mu_2\dots \mu_{p+1}}(\, \vec s \, )
	  {\delta\over \delta\sigma^{\mu\mu_2\dots \mu_{p+1} }(\, \vec  s\, )}
	  \ ,\qquad
	  y^{\mu_2\dots \mu_{p+1}}(\, \vec s \, )\equiv
	  \epsilon^{m_2\dots\mu_{p+1}}\, \partial_{m_2}\, y^{\mu_2}\dots
	  \partial_{m_{p+1}}\,  y^{\mu_{p+1}}\ .
	   \end{equation}

	   The above relationship can be used to describe the {\it shape
deformations}, or local distortions of the $p$--brane in terms of the
Jacobi equation
	 in $p$--\textit{loop space} \cite{jacobi}. The boundary effective
         action $S_B$   leads us to define the
	  {\it boundary momentum density} as
	  the dynamical variable canonically conjugated to the boundary
	  coordinate $y^\mu(\vec s)$:

	  \begin{equation}
	  {\delta S_B\over \delta y^\mu(\, \vec s\, )}\equiv P_\mu(\, \vec
s\,)=
	  {1\over p !}\, \epsilon^{m_2\dots m_{p+1}}\,
	   P^{0)}{}_{\mu\mu_2\dots\mu_{p+1}}\, \partial_{m_2}\, y^{\mu_2}\dots
	  \partial_{m_{p+1}}\, y^{\mu_{p+1}} +N^m \,\partial _m\, \phi_\mu\ .
	  \end{equation}

	Here, $P_\mu(\, \vec s\, )$ describes the overall response of the
	$p$--brane boundary to {\it local} volume deformations
	encoded into $d\sigma^{\mu_1\dots  \mu_{p+1} }(\, \vec s\, )$, as
        well as to
	induced harmonic deformations, orthogonal to the boundary, described
	by the normal derivative of $\phi_\mu$.
	In a similar way, we can define the {\it energy density} of the system
	as the dynamical variable canonically
	conjugated to the  $p$--brane history volume variation

	  \begin{equation}
	  {\partial S_J\over\partial \Omega_{p+1}} =
	{1\over 2 m_{p+1}\, (p+1) ! }\, P^{0)}{}_{\mu\mu_2\dots\mu_{p+1}}
	  P^{0)}{}^{ \mu\mu_2\dots\mu_{p+1}} -\frac{m_{p+1}}{2}\, p\ .
	  \end{equation}

	  Finally, from the anti-symmetry of
$P^{0)}{}_{\mu\mu_2\dots\mu_{p+1}}$ under
	  index permutations we deduce the following identity

	  \begin{equation}
	  P_\mu\, P^\mu \equiv
	  { h\over  (p+1) ! }\, P^{0)}{}_{\mu\mu_2\dots\mu_{p+1}}\,
	   P^{0)}{}^{ \mu\mu_2\dots\mu_{p+1}}\ .
	  \end{equation}

	  Thus, we arrive at the main result of the classical formulation
in the form of a {\it reparametrization invariant,
	 relativistic, effective Jacobi equation}

	 \begin{equation}
	  {1\over 2 m_{p+1}V_p}\,
	 \int_{\Sigma_{p}}{ d^{ p}s\over \sqrt h }\, \left(\,
	  { \delta S^{eff}\over \delta y_{\mu }(\, s\, )}-N^m(\, s\, )\,
	  \partial_m\, \phi_\mu\, \right)\,
	  \left(\, {\delta S^{eff}\over \delta y^\mu (\, s\, )}-N^j(\, s\, )\,
	  \partial_j\, \phi^\mu\, \right) -\frac{m_{p+1}}{2}\, p
	 = {\partial S^{eff}\over\partial \Omega_{p+1}}\ .
	 \label{jacreleq}
	 \end{equation}

	{\it This Jacobi equation encodes the boundary dynamics of the
	$p$--brane with respect
	 to an evolution parameter represented by the world--volume of the
	 $p$-brane history.} This is a generalization of the {\it areal} string
	 dynamics originally introduced by Eguchi \cite{eguchi}
	 via reparametrization of the Schild action \cite{schild},
	 \cite{sec2}.

\subsection{``Classical Quenching'' $\to$ Volume Dynamics}

The Jacobi equation derived in the previous subsection takes into account
both the intrinsic
	 fluctuations $\delta y^{\mu}(\, \vec s\, )$ and the normal  boundary
	 deformations $dN^m\, \partial_m\, \phi^\mu$ induced by the bulk field
	 $\phi^\mu$. The problem is that, even neglecting the
	 boundary fluctuations induced by the bulk harmonic mode,
	 the $p$--\textit{loop space} Jacobi equation is difficult to handle
	 \cite{noi2}, \cite{noi3}, \cite{noi4}. In order to make some
progress, it is necessary to forgo the {\it local} fluctuations of the
brane in favor of the simpler,
	{\it global} description in terms of hyper-volume variations,
	without reference to any specific point on the $p$-brane  where a
local fluctuation may actually
	occur.\\
Thus, in our formulation of $p$--brane dynamics, {\it classical quenching}
means having to relinquish the idea of describing the local deformations
$\phi^\mu
(\sigma)$ of the brane, and to focus instead on the collective
	mode of oscillation. In turn, by ``collective dynamics,'' we mean
	{\it volume variations} with no reference to the local fluctuations
which
       cause the volume to vary. In this approximation we can write a
``global'', i.e., non-functional, $p$--brane wave
	equation.\\

The effective action that encodes the volume
	dynamics, say $S _{0}$, is obtained from $S_J + S_B$ by ``freezing''
	both the harmonic and Kalb--Ramond bulk modes. The simplification
is that the general action reduces to the following form
	\begin{eqnarray}
	S_J+ S_B\longrightarrow   S_0 & = &
	{ 1\over (p+1)!}\, \sigma^{\mu\mu_2\dots\mu_{p+1}}\,
         P^{0)}{}_{\mu\mu_2\dots\mu_{p+1}}  \nonumber\\
	&& + \Omega_{p+1}\left[\,
	{ 1\over 2 m_{p+1}(p+1)!}\,  P^{0)}{}_{\mu\mu_2\dots\mu_{p+1}}
	  P^{0)}{}^{ \mu\mu_2\dots\mu_{p+1}} - \frac{m_{p+1}}{2}\, p\,
	\right]\,
	\label{s0}
	\end{eqnarray}

so that the functional equation reduces to a partial differential equation

	 \begin{equation}
	 {1\over 2 m_{p+1}}\,{ \partial
   S_0\over \partial \sigma_{\mu_1\dots\mu_{p+1} }}
   {\partial S_0\over \partial
	\sigma^{\mu_1\dots\mu_{p+1}}} -\frac{m_{p+1}}{2}\, p
	 = {\partial S_0\over\partial \Omega_{p+1}}\ .
	 \label{minijac}
	 \end{equation}

	 The collective--mode dynamics is much simpler to handle. In fact,
	 the functional derivatives that describe the shape variation of
the brane
	 have been replaced by ``ordinary'' partial derivatives that take into
	 account only hyper--volume variations, rather than local distortions.
	 In other words, while the original equation {  (\ref{jacreleq})}
	 describes the
	{\it shape dynamics}, the global equation  {   (\ref{minijac})}
	 accounts for the {\it collective dynamics} of the brane. The advantage
	 of the Jacobi equation  {  (\ref{minijac})} is that the partial
	 derivative is taken with respect to a {\it matrix coordinate}
	 $\sigma_{\mu_1\dots\mu_{p+1}}$ instead of the usual position
	four--vector.\\
The similarity with the point particle case $(p=0)$
	suggests the following ansatz for $S_0$

	 \begin{equation}
	 S_0\left( \, \sigma\ ;\Omega\, \right)\equiv
	 {B\over 2 \Omega_{p+1}}\,
	{1\over  (p+1)!}\,  \left(\, \sigma^{\mu_1\dots\mu_{p+1}}-
	\sigma^{\mu_1\dots\mu_{p+1}}_0\, \right)^2 -\frac{m_{p+1}}{2}\, p\,
	V_{p+1}\ ,
	 \end{equation}

where $\sigma^{\mu_1\dots\mu_{p+1}}_0$ represents a constant ( matrix ) of
integration
	 to be determined by the ``initial conditions'', while
	 the value of the $B$ factor is fixed by the equation {
	 (\ref{minijac})}. Indeed

	 \begin{equation}
	{ \partial S_0\over \partial \sigma^{\mu_1\dots\mu_{p+1} }}=
	{B\over \Omega_{p+1}}\, \left(\, \sigma_{\mu_1\dots\mu_{p+1}}-
	\sigma_{0)\mu_1\dots\mu_{p+1}}\, \right)
	 \end{equation}

         and

	 \begin{equation}
	{\partial S_0\over\partial \Omega_{p+1}}=
	 -{B\over 2\Omega_{p+1}\, (p+1)! }\left(\,
	\sigma_{\mu_1\dots\mu_{p+1}}-
	\sigma_{0)\mu_1\dots\mu_{p+1}}\right)^2-\frac{m_{p+1}}{2}\, p\,
	 \Omega_{p+1}\,
	 \end{equation}

     	so that

	  \begin{eqnarray}
	 && B= -m_{p+1}\,\\
	 && S_0\left(\sigma;V\right)=
	  - { m_{p+1} \over 2 V_{p+1}}\,
	{1\over  (p+1)!}\, \left(\, \sigma^{\mu_1\dots\mu_{p+1}}-
	\sigma^{\mu_1\dots\mu_{p+1}}_{0)}\, \right)^2 -\frac{m_{p+1}}{2}\,
	p\,  V_{p+1}\ .
	 \end{eqnarray}

	For the sake of simplicity, one may choose the integration constant
to vanish, i.e., one may set
	 $\sigma^{\mu_1\dots\mu_{p+1}}_{0)}=0$. Thus, in the quenching
approximation, we have obtained the classical Jacobi action for the
hyper-volume
	 dynamics of a free $p$-brane.

	 \section{The Full Quantum Propagator}

\subsection{``Momentum Space'' Propagator}

Eventually, one is interested in computing the quantum amplitude for the brane
	to evolve from an initial (vacuum) state to a final, finite
	volume, state. In general, the $p$--brane ``two--point'' Green
function represents the
	 correlation function between an initial brane configuration
	 $y_0(\vec s)$ and a final configuration   $y(\, \vec s\, )$. In
the {\it quantum theory of $p$--branes} the Green function is obtained as a
sum over
	 all possible histories of the world--manifold $\Sigma _{p+1}$
	in the corresponding phase space. For the sake of simplicity, one may
	 ``squeeze'' the initial boundary of the brane history
	 to a single point. In other words, the physical process that we
have in mind represents the quantum nucleation of a $p$--brane so that the
propagator that we wish to determine connects an initial brane of zero size
to a final object of proper volume
	 $V_p$. The corresponding amplitude $ G $ is represented by the
path--integral:

	 \begin{eqnarray}
	{\cal G} &\equiv&\int  \left[\, {\cal D}g_{m n}\, \right]
	 \int^x_{x_0} \left[\,  {\cal D} x(\tau)\, \right]
	 \int \left[\, {\cal D} p(\tau)\, \right]
	 \int^y \left[ \, {\cal D} Y^\mu\, \right]
	 \left[ \, {\cal D} P^m{}_\mu \, \right]
	 \times \nonumber\\
	&& \times \exp \left\{ \, i\,\int_{\Sigma _{p+1}} d^{p+1}\sigma \,
	\sqrt g\, \left[\,
	p_\mu\, \dot x^\mu - g^{00}{1\over 2 m_{p+1}}\, p_\mu \, p^\mu +
	{m_{p+1}\over 2 }\, \right]\right.
	\nonumber\\
	&& \quad -  i \left.\int_{\Sigma _{p+1}} d^{ p+1}\sigma\, \sqrt{g}\,
	\left[\,
	P^{ m}{}_\mu\, \nabla_m\,  Y^\mu(\, \sigma\, ) -{1\over 2 m_{p+1} }\,
	 g_{mn}\, P^{ m}{}_\mu \, P^\mu{} ^{ n} -p\, {m_{p+1}\over 2}\,
	 \right]\, \right\}\ .
	 \label{gfeyn}
	 \end{eqnarray}

	 Summing over ``all'' the brane histories in phase space means
	 summing over all the  dynamical variables, that is,
	  the {\it shapes} $Y^\mu(\, \sigma\, )$ of the
world--manifold
	 ${\Sigma _{p+1}}$, over the
	 {\it rates of shape change}, $P^m{}_\mu(\, \sigma\, )$, and over
	 the bulk {\it intrinsic geometries}, $g_{mn}(\, \sigma\, )$, with the
	 overall condition that the shape of the boundary is described by
	 $x^\mu=y^\mu(\vec s)$ and its intrinsic geometry by
	 $h_{ij}(\, \vec s\,)$. Thus,

         \begin{eqnarray}
	 \left[\, {\cal D} g_{m n}\,  \right] &\equiv &\left[\, Dg_{mn}\,
	 \right]\,\delta\left[ \, g_{mn} -\overline{g}_{mn}\,
\right]\nonumber\\
	&=& \left[ \, {\cal D}g_{00}\,  \right] \left[\, {\cal D} e\, \right]
	 \delta\left[ \, g_{00}-e^2(\tau) \, \right]
	 \left[  {\cal D}g_{ik}   \right]\, \delta \left[\,
	  g_{ik}-h_{ik}(\, \vec s\, )  \,\right]\ ,
	 \end{eqnarray}

where the only non-trivial integration is over the unconstrained
	 field $ e(\, \tau\, )$.\\
Carrying out all three functional integrations would
	 give us a {\it boundary} effective theory encoding all the information
	 about bulk quantum dynamics, in agreement with the {\it Holographic
	 Principle}. However, there are several technical difficulties that
need to be overcome before reaching that goal.\\
	Let us begin with the shape variables $Y^\mu(\, \sigma\, )$; they
appear in the path--integral as ``Fourier
	 integration variables'' linearly conjugated to the classical
	 equation of motion. Hence, the $Y$ integration gives a (functional)
	 Dirac--delta that confines $P^m{}_\mu$ on-shell:

	 \begin{equation}
	 \int[  {\cal D} Y^\mu]\exp\left\{-i \int_{\Sigma _{p+1}} d^{ p+1}
	 \sigma\, \sqrt{g} Y^\mu(\sigma)\nabla_m P^{ m}{}_\mu \right\}\propto
	 \delta\left[\nabla_m P^{ m}{}_\mu \right]\ .
	 \label{dclass}
	 \end{equation}

	The proportionality constant in front of the Dirac--delta is
	 physically irrelevant and can be set equal to unity.\\
	 Integrating out the brane coordinates is equivalent to ``shifting
	 from configuration space to momentum space'' in a functional
	 sense. However, the momentum integration is not free, but is
restricted by Eq.{  (\ref{dclass})} to the family of classical
	 trajectories that are solutions of equation {
(\ref{transv})}. Then, we can
	 write the two--points Green function as follows

	 \begin{equation}
	\overline G= N\,
	\int  \left[\, dP^{0)}\, \right] \int  \left[\, {\cal D}g_{m n}\,
	\right]\int  [\, {\cal D} \phi \, ][\, {\cal D} A \, ]\, \exp
	\left (\, i S^{eff}\, \right)\ ,
	 \end{equation}
	where $N$ is a normalization factor to be determined at the end of
	 the calculations, $S^{eff}$ is the effective action {(\ref{jeff})}
	 and we integrate over the zero mode components in the ordinary
sense, that is, we integrate over numbers and not over functions.
	 It may be worth emphasizing that  we have traded the
	 original set of scalar fields $Y^\mu(\sigma)$ with the ``Fourier
	 conjugated'' modes  $\phi $, $A$, $P^{0)}$. Then

	\begin{equation}
	\overline G= N\,
	 \int  [\, {\cal D}g \,]
	 \exp\left\{ -i\frac{m_{p+1}}{2}\, p\,
	 \int_{\Sigma _{p+1}} d^{ p+1}\sigma\, \sqrt{-\det g_{mn}}\, \right\}
	 K\left[\, \sigma\ ; g\, \right]\, Z _{\phi , A} \left[\, g\, \right]
	 \ .
	 \end{equation}

	 We recall that

	 \begin{equation}
	 V_{p+1}\equiv \int_{\Sigma _{p+1}} d^{ p+1}\sigma\,
	 \sqrt{-\det g_{mn}}=\int_{\Sigma _{p}} d^p s\,
	 \sqrt{-\det h_{ij}}\,\int_0^T d\tau \, e(\, \tau\, ) \equiv
	 V_p \int_0^T d\tau \, e(\, \tau\, )
	 \end{equation}

	 and

	 \begin{eqnarray}
	 K\left[\, \sigma\ ; e \, \right]=&& N\,
	 \int  \left[\, dP^{0)}_{\mu_1\dots \mu_{p+1}}\, \right]
	 \exp\left\{\, { i\over (p + 1) !}\,
	P^{0)}_{\mu\mu_2\dots\mu_{p+1}}\, \sigma^{ \mu\mu_2\dots\mu_{p+1}}
	 \, \right\}\times\nonumber\\
	 &&\exp\left\{\,  i\, V_p\,  {2m_{p+1}\over (p + 1)!}
	 \int_0^T d\tau\,  e(\, \tau\, )\,
	 P^{0)}_{\mu\mu_2\dots\mu_{p+1}}\, P^{0)\mu\mu_2\dots\mu_{p+1}}\,
	   \right\}\ .
	 \label{kbound}
	 \end{eqnarray}

	 At this point, we would like to factor out of the whole
	 path--integral the boundary dynamics, i.e. we would like to write
	 $ K = K(\, boundary\,) \times K(\, bulk\,)$. In order to achieve this
	 splitting between bulk and
	 boundary dynamics, we need to remove  the dependence on the
$00$--component of the bulk metric $g_{mn}$ in $ K\left[\, \sigma\ ; e\,
	 \right]$. In other words, we are looking for a propagator
where $\sqrt{\vert\sigma\vert }$ plays the role
	 of ``euclidean distance'' between the initial and final
	 configuration. In support of this interpretation, we also need a
suitable parameter that plays the role of ``proper time''
	 along the history of the branes connecting the initial and final
	 configurations. The obvious candidate for that role
	 is the {\it  proper time lapse} $\int_0^T d\tau\, e(\tau)$.
However, the $e$--field is subject to quantum fluctuations, so that the
proper time lapse is a quantum variable itself. Accordingly, a
	 $c$--number $\Omega_{p+1}$ can be defined {\it only} as a
	 {\it quantum average} of the proper world volume operator

	 \begin{equation}
	 \Omega_{p+1}= V_p\, \langle\,  \int_0^T d\tau\, e(\,\tau\,)\,
	 \rangle\ .
	 \end{equation}

	 Thus, we replace the quantum  proper volume $V_{p+1}[\,g\, ]$ with
the
	 quantum average $\Omega_{p+1}$ in $ K\left[\, \sigma\ ; e \, \right]$
	 and write the boundary propagator in the form

	 \begin{eqnarray}
	 K\left[\, \sigma\ ; \Omega_{p+1}\, \right] & = & N\,
	 \int  \left[\, dP^{0)}\, \right]\,
	 \exp\left\{ \, { i\over (p + 1) !}\,
	P^{0)}_{\mu\mu_2\dots\mu_{p+1}}\,
	 \sigma^{ \mu\mu_2\dots\mu_{p+1}}+ \right.
 	 \nonumber \\
	 & &
	 \qquad \left.
	 + i
	 \frac{\Omega_{p+1}}{2m_{p+1} (p + 1)!}\,
	 P^{0)}_{\mu\mu_2\dots\mu_{p+1}}\, P^{0)\mu\mu_2\dots\mu_{p+1}}\,
	 \right\}\ ,
	 \label{kbound2}
	 \end{eqnarray}

	 while the amplitude becomes

	 \begin{eqnarray}
	 \overline G=&& N \, \int_0^\infty d\Omega_{p+1}\,
	  \exp\left\{ -i\frac{m_{p+1}}{2}\,  p\, \Omega_{p+1}\, \right\}
	  K\left[\, \sigma\ ;\Omega_{p+1}\, \right]
	  \nonumber\\
	 &&\times\int [\, {\cal D}g\, ]\, \delta\left[\, \Omega_{p+1}- V_p\,
	\langle\, \int_0^T \, d\tau\,  e(\, \tau\, ) \,
	 \rangle \, \right]\,  Z_{\phi, A}\left[\,  g\, \right]\label{green}\ .
	 \end{eqnarray}

	The ``bulk'' quantum physics is encoded now into the path--integral

	 \begin{eqnarray}
	 Z_{\phi , A}\left[\, g\, \right] &=&
	 \int[\, {\cal D} \phi\, ] [ \, {\cal D} A \, ]
	\exp \left\{ \, {i\over 2m_{p+1}(p+1)!}
        \right.
	\nonumber\\
	& & \times
	\left. \int_{\Sigma _{p+1}} d^{ p+1}\sigma\,
	\sqrt{g}\, \left(\,
	g^{mn} \, \partial_m\, \phi_\mu \,  \partial_n\, \phi^\mu -
	{1\over p!}\, F^\mu{}_{m_1\dots m_p}(\, A\, )\,
	F_\mu{}_{m_1\dots m_p}(\, A\, )\, \right)\,
	 \right\}
         \ .
	  \end{eqnarray}

	Equation {  (\ref{green})} presents a new problem: the constraint
	over the metric integration, which allowed us to factor out the
boundary
	dynamics, is highly non-linear as it depends on the vacuum average
	of the quantum volume. However, we can get around this difficulty
by replacing the Dirac delta with an exponential weight factor

	  \begin{equation}
	 \delta\left[\, \Omega_{p+1}- V_p\, \langle\,  \int_0^T d\tau\,
	 e(\, \tau\, )\, \rangle \, \right]
	 \quad \longrightarrow \quad \exp\left\{- i\, \Lambda\,
	 \left(\, \Omega_{p+1}-V_p \, \int_0^T d\tau \, e(\, \tau\, )\,
	 \right)\, \right\}\ ,
	 \end{equation}

	where $\Lambda$ is a {\it constant Lagrange multiplier}. Thus,
we first perform all calculations with $\Omega_{p+1}$ as an arbitrary
	evolution parameter and only at the end we impose the condition

	 \begin{equation}
	 {\partial \over\partial\,\Lambda}
	 \overline G =0 \quad\Longrightarrow\quad \Omega_{p+1}=
	 V_p \, \langle \, \int_0^T d\tau \, e(\, \tau\, )\, \rangle
         \ .
	 \label{onshell}
	 \end{equation}

	 In this way, we can write the $\Lambda$ dependent  amplitude in
the form

	 \begin{equation}
	\overline G =
	 N \,\int_0^\infty d\Omega_{p+1}\,
	\exp\left\{ -i \Omega_{p+1}\left(\, \Lambda +{m_{p+1}\over 2}\,
	p\, \right)\,
	\right\}\, K\left[\, \sigma \ ;\Omega_{p+1}\, \right]\times
	\int [\, {\cal D}g \, ]\, Z_{\phi A}\left[\, g\ ;\Lambda \,\right]
         \ ,
\end{equation}

where the bulk quantum physics is encoded into the path--integral

	 \begin{eqnarray}
	 Z_{\phi , A}\left[\, g\ ;\Lambda\, \right]&=&\exp \left\{i\,
	\Lambda \, V_p\,
	 \int_0^T\, d\tau\, e(\, \tau\, )\, \right\}\, \int[\,{\cal D}\phi\,]
	 [\,{\cal D} A\,]\,\exp \left\{\, {i\over 2 m_{p+1}(p+1)!} \right.
        \nonumber\\
	 && \times\left.
	 \int_{\Sigma _{p+1}} d^{ p+1}\sigma\, \sqrt{g}\, \left(\,
	  g^{mn}\, \partial_m\, \phi^\mu \, \partial_n\, \phi_\mu -
	  {1\over p!}\, F^\mu{}_{m_1\dots m_p}(\, A \,)\,
	 F_\mu{}_{m_1\dots m_p}(\, A\, )\, \right)\,
	 \right\}
\end{eqnarray}

\subsection{The Boundary Propagator}

The main point of the whole discussion in the previous subsection is this:
{\it even though the bulk quantum dynamics cannot be solved exactly, since
there is no way  to compute the bulk fluctuations in closed form, the
boundary propagator can be evaluated exactly.} This is because the integral
{  (\ref{kbound2})} is gaussian in $P^{0)}$:

	 \begin{equation}
	 K\left[\, \sigma\ ;\Omega_{p+1}\, \right]=\left[\,
	 {m_{p+1}\over i\pi \Omega_{p+1}}\, \right]^{{1\over 2}{D\choose p+1}}
	 \exp\left\{\,  {i m_{p+1}\over 2(p+1)! \Omega_{p+1}}\,
	\sigma^{\mu_1\dots\mu_{p+1}} \, \sigma_{\mu_1\dots\mu_{p+1}}\,
	   \right\}\label{prop0}\ .
\end{equation}

Moreover, one can check through an explicit calculation that the
	 kernel $K$ solves the (matrix) Schroedinger equation

	 \begin{equation}
	 \left[\, {1\over 2m_{p+1}  (p+1) !}\,
	   {\partial^2 \over
	 \partial \sigma^{ \mu_1\dots \mu_{p+1}}\,
	 \partial\sigma_{\mu_1\dots\mu_{p+1}}}\, \right]\,
	 K\left[ \, \sigma-\sigma_0\ ; \Omega_{p+1}\, \right] =
	 -i{\partial \over \partial \Omega_{p+1}}\,
	 K\left[ \, \sigma-\sigma_0\ ;\Omega_{p+1}\, \right]
	 \label{schroed}
	 \end{equation}

with the boundary condition

	 \begin{equation}
	 \lim_{\Omega\to 0} K\left[\,\sigma-\sigma_0\ ;\Omega_{p+1} \,\right]
	 =\delta\left[\, \vert \sigma-\sigma_0\vert \, \right]
	 \label{bc}\ .
	 \end{equation}

	Notice that the average proper volume $\Omega_{p+1}$
	 enters the expression of the kernel $K$ only through the
combination $m_{p+1}/\Omega_{p+1}$.
	 Thus, the limit {  (\ref{bc})} is physically equivalent to the
	 {\it infinite tension limit} where $\Omega_{p+1}$ is kept fixed
	 and $m_{p+1}\to\infty$:

	 \begin{equation}
	 \lim_{m_{p+1}\to \infty} K\left[ \, \sigma-\sigma_0\ ;\Omega_{p+1}
\,\right]
	 =\delta\left[ \, \vert \sigma-\sigma_0\vert \, \right]\ .
	 \label{point}
	 \end{equation}

	 In the limit {  (\ref{point})} the infinite tension shrinks the
	 brane to a pointlike object.\\
	 From the above discussion we infer that the quantum dynamics
	of the collective mode can be described either by the zero mode
	propagator {  (\ref{prop0})}, or by the wavelike equation

	 \begin{eqnarray}
	&& \left[ \, {1\over 2m_{p+1}  (p+1) !}\,
	   {\partial^2 \over
	 \partial \sigma^{ \mu_1\dots \mu_{p+1}}\, \partial\sigma_{\mu_1\dots
	\mu_{p+1}}}\, \right]\, \Psi_0\left[\,  \sigma\ ;\Omega_{p+1}\,
\right]
	 = -i{\partial \over \partial \Omega_{p+1}}
	 \!\Psi_0\left[ \,  \sigma;\Omega_{p+1}\,  \right]
         \nonumber \\
	 && \Psi_0\left[ \,\sigma\ ;\Omega_{p+1}\, \right]=\int \!\!\left[\,
	 d\sigma_0\, \right]\,  K\left[ \,  \sigma-\sigma_0\ ;\Omega_{p+1}\,
	 \right]\, \phi\left[\, \sigma_0\ ;0\, \right]\ ,
	 \end{eqnarray}

where $\phi\left[ \, \sigma_0\ ;0 \, \right]$ represents the initial
	 state wave function. Comparing the ``Schroedinger equation''
	 {  (\ref{schroed})} with the Jacobi equation
	 {  (\ref{jacreleq})} suggests the following
	 {\it Correspondence Principle} among classical variables and quantum
	 operators:

	 \begin{eqnarray}
	 P^{0)}_{\mu_1\dots\mu_{p+1}} &\quad\longrightarrow\quad& i\,
	 {\partial\over \partial\sigma^{\mu_1\dots\mu_{p+1}}}
	 \ ,\qquad\left(\, p\ge 1\,\right)
	 \\
	  E &\quad\longrightarrow\quad& -i{\partial\over\partial \Omega_{p+1}}
         \ .
	 \end{eqnarray}
	In summary, the main result of this section is that the
general form of the quantum propagator for a closed
	 bosonic $p$-brane can be written in the following form
	 \begin{eqnarray}
	\overline G &=&N\int_0^\infty d \Omega_{p+1}\, \exp\left\{\,  i
	  \Omega_{p+1}\left[\,\Lambda + {m_{p+1}\over 2} \, p \, \right]\,
	  \right\}\, \left[\, {m_{p+1}\over i\pi \Omega_{p+1}}\,
	  \right]^{{1\over 2}{D\choose p+1}}\nonumber\\
	 &&\times\exp\left\{\, {i m_{p+1}\over 2  (p+1)!\Omega_{p+1}}\,
	 \sigma^{\mu_1\dots\mu_{p+1}} \, \sigma_{\mu_1\dots\mu_{p+1}}\,
	\right\}
	 \int \left[\, {\cal D}e\, \right] K_{cm}\left[\,  x-x_0\ ;
	 e(\, \tau\, )\, \right]\, Z\left[\,  e\ ;\Lambda\, \right]
        \label{what}\ .
	 \end{eqnarray}

 \section{``Minisuperspace--Quenched Propagator''}

At this stage in our discussion, we can explicitly define our approximation
scheme. It consists of three main steps.\\
The first essential step, discussed in subsection IIB, consists in
splitting the metric of the world--manifold according to Eq.(\ref{ds1}). In
a broad sense, this is a ``minisuperspace approximation'' to the extent
that it restricts the general covariance of the action in parameter space.
In other words, separating the center of mass proper time from the spatial
coordinates on the world manifold $\Sigma_{p+1}$ effectively breaks the
	full invariance under general coordinate transformations into two
symmetry groups:

	\begin{equation}
	\mathrm{General\quad Diffs} \longrightarrow (\mathrm{time})\mathrm{Rep}
	\otimes
	(\mathrm{spatial})\mathrm{Diffs}\ ,
	\end{equation}

so that the metric (\ref{ds1}) shows a residual symmetry under
	independent time reparametrizations and spatial diffeomorphisms. By
virtue of this operation, we were able to separate the center of mass
motion from the bulk and boundary dynamics. This is encoded in the split
form (\ref{ssplit}) of the total action for the $p$--brane.\\
The second and more strict interpretation of the minisuperspace
approximation is that we  now ``freeze'' the world
	metric into a background configuration $\overline{g}_{mn}$ where
	the space is a $p$-sphere, while
	$g_{00}$, or its square root  $e(\, \tau\, )$, is free to
	 fluctuate. In other words, we work in a
	{\it minisuperspace} of all possible world geometries.\\
	The third and last step in our procedure is an adaptation of  one
of the most useful approximations to the
	exact dynamics of interacting quarks and gluons, namely, ``{\it
Quenched}
	QCD''. There, the contribution of the determinant of the quark kinetic
	operator is set equal to one. In the same spirit we ``quench''
	all the {\it bulk} oscillations:

	 \begin{equation}
	 Z_{\phi A}\left[\,  e\ ;\Lambda\, \right]\longrightarrow
	 \exp\left\{-i\, \Lambda\, \Omega_{p+1}\, \right\}\ .
	 \end{equation}

        Combining these three steps, we obtain from (\ref{what})
	the {\it quenched--minisuperspace propagator}:

	 \begin{eqnarray}
	G\left[\, x-x_0\ , \sigma\, \right]
	  &=&  N\, \int_0^\infty d \Omega_{p+1}\,
	  \exp\left\{ \, i\, p\, \Omega_{p+1}{m_{p+1}\over 2}\, \right\}
	 \left[\, {m_{p+1}\over i\, \pi\, \Omega_{p+1}}
	 \right]^{{1\over 2}{D\choose p+1}}\times\nonumber\\
	 &&\exp\left\{ \, {i\,  m_{p+1}\over 2 \, \Omega_{p+1}}{\sigma^2
	 \over (p+1)!}\, \right\}
	 \int \!\!\left[\, {\cal D}e\, \right]\, K_{cm}\left[\, x-x_0\ ;
	 e(\, \tau\, )\, \right]\, \delta\left[\, \int_0^T \!\!d\tau\,
	 e(\, \tau\, )-{\Omega_{p+1}\over V_p}\, \right]\ .
	 \end{eqnarray}

The center of mass propagator $ K_{cm}\left[\, x-x_0\ ;
	 e(\, \tau\, )\, \right]$ can be computed as follows \cite{nrprop}.
	From the  Lagrangian $L_{cm}$ we can define the
	 {\it center of mass momentum} $p_\mu$ as follows

	 \begin{equation}
	 p_\mu\equiv {\partial L_{cm}\over \partial\dot x^\mu(\, \tau\, ) }=
	 M_0\, {\dot x_\mu \over e(\, \tau\, )}\ .
	 \end{equation}

	 By Legendre transforming $L_{cm}$ we obtain the {\it center of mass
	 Hamiltonian}:

	 \begin{equation}
	 H_{cm}\equiv p_\mu\, \dot x^\mu -L_{cm} ={e(\tau)\over 2M_0}\left[\,
	 p_\mu \, p^\mu  + M_0^2\, \right]\ .
	 \end{equation}

Moreover, the canonical form of $S_{cm}$ reads

	 \begin{equation}
	 S_{cm}=  \int _0^T d\tau\, \left[\,  p_\mu\, \dot x^\mu
	-{e(\, \tau\, )\over 2M_0} \,  \left(\,
	 p_\mu \,  p^\mu + M_0^2\,  \right)\, \right]\,
	 \end{equation}

so that the quantum dynamics of the bulk center of mass is described
	 by the path--integral

	 \begin{equation}
	 K_{cm}\left[\, x-x_0\ ; e(\, \tau\, )\, \right]\equiv \int
	 \left[\, {\cal D }x\, \right]\, \left[\, {\cal D }p\, \right]\,
	 e^{i S_{cm}\left[\,  x\ , T\ ; e(\, \tau\, )\, \right]}\ .
	 \end{equation}

	 This path--integral can be reduced to an ordinary integral over
	 the constant four momentum $q_\mu$ of a point particle of mass $M_0$

	 \begin{equation}
	  K_{cm}\left[\, x-x_0\ ; e(\, \tau\, )\, \right]=
	  \int { d^D q\over (\, 2\pi\, )^D}\,
	  e^{i\, q_\mu\,  (\, x^\mu-x^\mu_0\, ) }\,
	  \exp\left[ - \int _0^T d\tau\, { e(\, \tau\, )\over 2M_0}\, \left(
	 \, q_\mu\,   q^\mu + M_0^2\,  \right)\, \right]\ .
	  \end{equation}
	In order to get the explicit form of the center of mass propagator we
	have to integrate, in the ordinary sense, over $q_\mu$. However,
	before that we must integrate, in the functional sense, over
	the einbein field $e(\, \tau\, )$.

	Using the properties of the Dirac--delta distribution, the einbein
field can be
	integrated out :

	 \begin{eqnarray}
	 \int \left[\, {\cal D }e\, \right]\delta\left[\,
	\int_0^T d\tau\, e(\, \tau\, ) - \Omega_{p+1}/V_p\, \right]
	&&\exp\left[
	 - i\int _0^T d\tau\,  {e(\, \tau\, )\over 2M_0}\,
	 \left(\,  q_\mu\, q^\mu + M_0^2\, \right)\, \right]=
	 \nonumber\\
	 &&\exp\left[ - i{\Omega_{p+1}\over 2M_0 V_p}\,
	 \left(\, q_\mu\,  q^\mu +\, M_0^2\, \right)\, \right]\,
	 \end{eqnarray}

	 and

	 \begin{equation}
	 \int { d^D q\over (2\pi)^D }\, e^{i\, q_\mu \, (\, x^\mu-x^\mu_0\, )}
	 \exp\left[ -i {\Omega_{p+1}\over 2M_0V_p }\, q_\mu\,  q^\mu\,
	 \right]=\left(\, {\pi M_0 V_p\over \Omega_{p+1} }\, \right)^{D/2}
	 \exp\left[ - i\, {M_0  V_p\over 2\Omega_{p+1} }\,\left(\, x -
	 x_0\right)^2\, \right]\ .
	 \end{equation}

         Using the above results, the ``QCD--QC combined approximation''
leads to the following
	expression for the propagator

	 \begin{eqnarray}
	G\left[\, x- x_0\ ,\sigma\, \right]
	  =&& N\, \int_0^\infty  d\Omega_{p+1}\,
	  \exp\left\{ \,  i\, \Omega_{p+1}\left[\,
	   {m_{p+1}\over 2}\, p + { M_0\over 2\,V_p}\,\right]\, \right\}
	 \left[\, {m_{p+1}\over i\pi \Omega_{p+1}}\right]^{{1\over 2}{D\choose
	 p+1}}\nonumber\\
	 &&\times \exp\left\{ \, {i m_{p+1}\over 2\Omega_{p+1}}
	{\sigma^2\over (p+1)!}\, \right\}\,
	 \left(\, {\pi M_0 V_p\over i\Omega_{p+1}}\, \right)^{D/2}
	 \exp\left[\,  i{M_0  V_p\over 2\Omega_{p+1} }\, \left(\, x -
	 x_0\, \right)^2 \, \right]\ .
	 \label{minik}
	 \end{eqnarray}

	 The amplitude { (\ref{minik})} can be cast in a more familiar
form in terms of the Schwinger--Feynman parametrization

	 \begin{equation}
	 s\equiv {\Omega_{p+1}\over 4  V_p}, \quad d\Omega_{p+1}= 4V_p\,
	  ds\ .
	 \end{equation}

     Using the above parametrization, the quantum propagator takes its
final form in the minisuperspace--quenched approximation

	 \begin{eqnarray}
	G\left[\, x- x_0\ ,\sigma\ ; M_0\, \right]
	  =&& N\, V_p \int_0^\infty  ds\,
	  \left(\, {\pi M_0 \over is}\, \right)^{D/2}
	  \exp\left[\,  i {M_0  \over 2s }\left(\,  x -
	 x_0\, \right)^2 \, \right]\times\nonumber\\
	  &&\exp\left\{\,i\, s {M_0\over 2}\, (p+1)\,  \right\}
	 \left[\, {M_0\over i\pi V_p^2 s }\, \right]^{{1\over 2}{D\choose
	 p+1}}
	 \exp\left\{ \, {i M_0\over 2s V_p^2}{\sigma^2\over(p+1)!}\, \right\}
	 \nonumber\\
	  =&&{i\over 2M_0}\int_0^\infty  ds\,
	  \left(\, {\pi M_0 \over is}\, \right)^{D/2}\left[\, {M_0\over i\pi
	  V_p^2 s }\, \right]^{{1\over 2}{D\choose p+1}}\exp\left\{ \, i\, s
	 \, {M_0\over 2}\, (p+1)\, \right\} \times\nonumber\\
	 &&\exp\left\{\, i\, {M_0  \over 2s }\,\left[\, \left(\, x -
	 x_0\, \right)^2 + {1\over  V_p^2}{\, \sigma^2\over (p+1)!}\,\right]
	 \, \right\}\ .
\label{minis}
	 \end{eqnarray}

	Here we have set $N=i/2M_0\, V_p$ in order to match the form of the
point particle propagator.\\
The formula (\ref{minis}) represents the main result of all previous
calculations and holds for any $p$ in any number of spacetime dimensions.

\subsection{Green Function Equation $\to$ Tension--Shell Condition}

For completeness of exposition, in this subsection we derive the master
equation satisfied by the Green function (\ref{minis}) in the
quenched--minisuperspace approximation. Then, by formally inverting that
equation we arrive at an alternative expression for the Green function in
momentum space. The advantage of this procedure is that it provides a
useful insight into the structure of the $p$--brane propagator.\\
	 To begin with, it seems useful to remark that the propagation kernel
in {
	 (\ref{minis})} is the product of the center of mass kernel
	 $K_{cm}\left(\, x-x_0\ ; s\, \right)$ and the volume kernel
	 $K\left(\, \sigma\ ; s\,\right)$; each term carries a weight given
by the phase factor $\exp i\, M_0\, s $ and
	 $\exp i\,p\, M_0\, s $, respectively.
	 Finally, we integrate over all the values of the Feynman parameter
$s$:

	\begin{equation}
	G\left[\, x- x_0\ ; M_0\, \right]=  {i\over 2M_0}\, \int_0^\infty ds
	\exp\left\{\, i\, {M_0\over 2}\, s\, (p+1)\, \right\}\,
	K_{cm}\left(\,  x-x_0\ ; s\, \right)\, K\left(\, \sigma\ ; s\,
\right)\ .
	\end{equation}

	As is customary in the Green function technique, we may add an
infinitesimal imaginary part to the mass in the exponent,
	that is, $ (M_0/2)\to (M_0/2) +i\epsilon$, so that the oscillatory
phase turns into
	an exponentially damped factor enforcing convergence at the upper
	integration limit. The ``$i\epsilon$'' prescription in the exponent
	allows one to perform an integration by parts leading to the
following expression

	\begin{eqnarray}
	&&G\left[\, x- x_0\ ; M_0\, \right]
	= {1\over  M_0^2\, (p+1)}\, \left[\,  \exp\left\{\, i\,(\,
	{M_0\over 2} +i\epsilon\,) \, s
        \, (p+1)\, \right\} \, K_{cm}\left(\,  x-x_0\ ; s\, \right)\,
K\left(\,
        \sigma\ ;s\, \right)\, \right]^\infty_0\nonumber\\
	&&- {1\over \, M_0^2\, (p+1)}\,
	\int_0^\infty ds \, \exp\left\{ \, i\, (\, {M_0\over 2} + i\epsilon\,)
	\, s\, (p+1)\, \right\}\, {\partial\over\partial s}\,
	\left(\, K_{cm}\left(\, x-x_0\ ; s\, \right)\,
	K\left( \, \sigma\ ;s\, \right)\, \right)\ .
	\end{eqnarray}

	Convergence of the integral enables us to express the partial
derivative
	$\partial/\partial s$ by means of the diffusion equations
	for $K_{cm}$ and $K$ and to move the differential operators
$\partial_\mu
	\, \partial^\mu$, $\partial^2/\partial\sigma^2$ out of the integral:

	\begin{eqnarray}
	G\left[\, x- x_0\ ; M_0\, \right]
	&=& {1\over \, M_0^2\, (p+1)}\left[\,
        \delta\left(\,  x-x_0\, \right)\, \delta\left(\,
        \sigma\ ;s\, \right)\, \right]
        - {1\over M_0^3\, (p+1)}
        \int_0^\infty ds\, \exp\left\{ \,  i\, {M_0\over 2}\, s\, (p+1)\,
        \right\}\nonumber\\
        &&\times \left[\,
	  \left(\, K\left(\, \sigma\ ;s\, \right)\,
	\partial^\mu\,\partial_\mu\,
	K_{cm}\left(\,  x-x_0\ ; s\, \right)\, \right) \right.
	\nonumber\\
	  &&\qquad \left. \left(\,
	K_{cm}\left(\,  x-x_0\ ;s\, \right){V_p\over (p+1)!}\,
	{\partial^2\, K\left(\, \sigma\ ; s\, \right)
	\over\partial\sigma^{\mu_1\dots\mu_{p+1}}\,
	\partial\sigma_{\mu_1\dots\mu_{p+1}}}\right)\, \right]\nonumber\\
	&=&-{1\over  M_0^2\, (p+1)}\, \delta\left(\, x-x_0\, \right)\,
	\delta\left(\, \sigma\ ;s\, \right)\nonumber\\
	&+&{1\over M_0^2(p+1)}\,
	\left(\, \partial_\mu\,\partial^\mu +
	{V_p\over (p+1)!}\, {\partial^2\over
	 \partial\sigma _{\mu_1\dots\mu_{p+1}}\,
	   \partial\sigma ^{\mu_1\dots\mu_{p+1}}}
	\, \right)\, G\left(\, x-x_0\ , \sigma\ ; s\, \right),
	\end{eqnarray}

	from which we deduce the desired result,

	\begin{equation}
	\left[\,  \partial_\mu\, \partial^\mu + {V_p\over (p+1)!}\,
	{\partial^2\over \partial\sigma_{\mu_1\dots\mu_{p+1}}\,
         \partial\sigma^{\mu_1\dots\mu_{p+1}}}
        - (p+1)\, M_0^2\, \right]\,  G\left(\, x-x_0\ , \sigma\, \right)=
        \delta^D\left(\, x-x_0\, \right)\,
	\delta\left(\, \sigma\, \right)\ .
	\end{equation}

	This is the Green function equation for the non--standard
	differential operator $\partial^\mu\partial_\mu +V_p\,
	\partial^2/\partial\sigma^2$. Finally, we ``Fourier transform'' the
	Green function
	by extending the momentum space to a larger space that includes the
	volume momentum as well:

	 \begin{eqnarray}
	G(\,  x-x_0\ , \sigma\, )&=&\int {d^Dq\over (2\pi)^D}\int
	[\, dk_{\mu_1\dots\mu_{p+1}}\, ]
	\exp\left( \, {i\, q_\mu\, (\, x-x_0\, )^\mu +{i\over (p+1)!}\,
	k_{\mu_1\dots\mu_{p+1}}\,
	\sigma^{\mu_1\dots\mu_{p+1}} }\, \right)\nonumber\\
	&&\times {1\over q^2 + {V_p^2\over (p+1)!}\,
	k^2_{\mu_1\dots\mu_{p+1}} +(p+1)\, M_0^2}\ .
	\label{gkk}
	\end{eqnarray}

	The vanishing of the denominator in {  (\ref{gkk})} defines a new
	{\it tension--shell} condition:

	\begin{equation}
	q^2 + {V_p^2\over (p+1)!}
	k^2_{\mu_1\dots\mu_{p+1}} +(p+1)\, M_0^2=0\ .
	\label{disp}
	\end{equation}

	Real branes, as opposed to {\it virtual} branes, must satisfy the
condition {  (\ref{disp})} which links together center
	of mass and volume momentum squared. Equation (\ref{disp})
represents an extension of the familiar Klein--Gordon condition for
point--particles to relativistic extended objects.\\
Some physical consequences of the ``tension--shell condition'', as
well as the mathematical structure of the underlying spacetime geometry
\cite{pezz},
are currently under investigation and will be reported in a forthcoming
letter. In the next subsection we limit ourselves to check the consistency
of our results against some familiar cases of physical interest.

	\subsection{Checks}

	\subsubsection{Infinite Tension Limit}

When probed at low energy (resolution) an extended object effectively
	 looks like a {\it point--particle.} In this case,
	 ``low energy'' means an energy which is small compared with the energy
	 scale determined by the brane tension. In natural units, the
tension of a $p$--brane
	 has dimension: $\left[\, T_p\,\right]=(\, energy\,)^
	 {p+1}$. Thus, when probing the brane at energy $E<< (\,T_p\,)^{1/p+1}$
	 one cannot resolve the extended structure of the object. From this
	 perspective, the ``{\it point--like limit}'' of a $p$--brane is
	 equivalent to the ``{\it infinite tension limit}''. In either
         case, no
	 higher vibration modes are excited and one expects the brane
	 to appear concentrated, or ``collapsed'', in its own center of
          mass. This critical
	 limit can be obtained from the general result {
	 (\ref{minis})} by setting
	 $p=0$ and performing the limit $ V_p\to 0 $ using the familiar
	 representation of the Dirac--delta distribution:

	 \begin{equation}
	\delta(x)\equiv \lim_{\epsilon \to 0}\left(\,{1\over \pi\epsilon}\,
	\right)^{d/2}\exp \left(\, -x^2/\epsilon\,\right)\ .
	\end{equation}

	In our case: $x^2\rightarrow \sigma^2/(p+1)!$, $ \epsilon \rightarrow
	-i M_0/4s V_p^2$, $d\rightarrow{D\choose p+1}$, and the whole
dependence on the volume coordinates of the brane reduces
        to a delta function which is different from zero only when
        $\sigma = 0 $. In this case, $G\left[\, x- x_0\ ; M_0\, \right]$
        reduces to the familiar expression for the Feynman propagator for a
point particle of mass $M_0$,\\

	\begin{eqnarray}
	G\left[\, x- x_0\ ; M_0\, \right]
	  &=& \delta\left[\, \sigma^2\, \right]\, {i\over
2M_0}\,\int_0^\infty
	  ds\,\left( \, {\pi M_0 \over is}\, \right)^{D/2}
	  \exp\left( \, -i\,{ M_0\over 2}\,  s\, \right)
	 \exp\left\{\, i\, {M_0  \over 2s }\, \left(\, x - x_0\, \right)^2\,
	 \right\}\ .
	 \label{minis2}
       \end{eqnarray}

\subsubsection{Spherical Membrane Wave Function}

>From the results of the previous subsections we can immediately extract the
generalized Klein--Gordon equation for a $2$--brane in four spacetime
dimensions:

\begin{equation}
	\left[ \, \partial_\mu\, \partial^\mu + {V_2\over 3!}\,
	{\partial^2\over \partial\sigma_{\mu_1\mu_2\mu_3}\,
         \partial\sigma^{\mu_1\mu_2\mu_3 }}
        - (p+1)\, M_0^2\, \right]\, \Psi\left(\,  x\ , \sigma\, \right)=0\ .
        \end{equation}

	In the following we show briefly how this wave equation specializes
to the case of a gauge fixed, or spherical, membrane of fixed radius
\cite{ctuck}, \cite{minib}. Moreover, since it is widely believed that
$p$--brane physics may be especially relevant at Planckian energy, we
assume that our $2$--brane is a fundamental object characterized by Planck
units of tension and length. Thus,
	for a spherical $2$--brane of radius $R$ we have:

	\begin{equation}
	V_2=4\pi l_{\mathrm{Pl}}^2,
	\end{equation}

	\begin{equation}
	\sigma^{\mu_1\mu_2\mu_3 }=\delta^{[\, \mu_1  }_x \, \delta^{\mu_2  }_y
	    \, \delta^{\mu_3 \, ] }_z \, { 4\pi\over 3}\,  R^3,
	\end{equation}

	  \begin{equation}
	    d\sigma^{\mu_1\mu_2\mu_3 }=\delta^{[\, \mu_1  }_x \,
	    \delta^{\mu_2  }_y\,
	     \delta^{\mu_3 ] }_z \,  4\pi R^2\, dR,
	      \end{equation}

	      \begin{equation}
	      {\partial\over\partial\sigma^{\mu_1\mu_2\mu_3 }}=
	      {1\over 4\pi R^2}
	      \delta_{[\, \mu_1  }^x\, \delta_{\mu_2  }^y\, \delta_{\mu_3\,]}^z
	      \, {\partial\over\partial R}.
	      \end{equation}

	 Since there is no mixing between ordinary and volume derivative,
	  we can use the method of separation of variables to factorize
	  the dependence of the wave function on $x$ and $\sigma$

	      \begin{equation}
	      \Psi\left(\,  x\ , \sigma\, \right)=\phi(x)\, \psi_0\left(\,
	      R\, \right) \quad \Longrightarrow \quad \qquad
	      \left[\, \partial_\mu\, \partial^\mu  -  M_0^2\, \right]\,\phi(x)
	      =0\ .
	      \end{equation}

	Here, $\phi_0( x )$ represents the center of mass wave function, while
	   the ``relative motion wave function'' $\psi\left(\, R\, \right)$
	   must satisfy the following equation

	   \begin{equation}
	\left[\,  {1\over 4\pi R^2}\,
	{\partial\over \partial R   }\, {1\over  R^2}\, {\partial\over \partial
	R } - {p M_0^2\over  l_{Pl}^2}\, \right]\,  \psi\left(\, R\, \right)=0
        \ .\label{radial}
 	\end{equation}
 	Apart from some ordering ambiguities, equation (\ref{radial})
 	 is the wave equation found in Ref.
 	\cite{ctuck},\cite{minib}, \cite{minib2}, for the zero energy
 	eigenstate, provided we identify the membrane tension $\rho$
through the expression

 	\begin{equation}
 	\rho^2 ={ p\,  M_0^2\over 4\pi l_{\mathrm{Pl}}^2}\ .
 	\end{equation}

 	It may be worth emphasizing that our approach preserves time
reparametrization invariance throughout all computational steps, while the
conventional minisuperspace
 	approximation assumes a gauge choice from the very beginning.

\section{Aknowledgements}
The authors are indebted to Dr. Kai Lam for proofreading the manuscript.


\begin{thebibliography}{99}
	\bibitem{wdw}
	J.A. Wheeler, ``{\it Geometrodynamics and the Issue of the Final
	State, }'' in C.DeWitt and B.S.DeWitt ed., ``{\it Relativity
	Group and Topology,}'' Gordon and Breach 1964;\\
	J.A. Wheeler, ``{\it Superspace and the Nature of Quantum
	Geometrodynamics,}'' in C.DeWitt and B.S.DeWitt ed., ``{\it Batelle
	Rencontres: 1967 Lectures in Mathematics and Physics,}''
	Benjamin, NY, 1968;\\
	B.S. DeWitt, Phys.Rev. {\bf 160}, 1113, (1967).
	\bibitem{padma} J.V. Narlikar, T.Padmanabhan,
	``{\it Gravity, Gauge Theories and Quantum Cosmology,}''
	D.Reidel Publ. Co., 1986.
	\bibitem{hh} J.B. Hartle, S.W. Hawking,
	Phys.Rev.D {\bf  28}, 2960 (1983).
	\bibitem{gianni} R. Casadio, G. Venturi,
	Class.Quant.Grav.{\bf 13},2375,  (1996);\\
	 G. Venturi,
	 Class.Quant.Grav.{\bf 7}, 1075,  (1990).
	\bibitem{quench} H.Hamber, G. Parisi, Phys.Rev. Lett. {\bf 47},
	1792, (1981); E.Marinari, G.Parisi, C.Rebbi, Phys. Lett. {\bf 47},
	1795, (1981); D.H.Weingarten, Phys. Lett. {\bf B109}, 57, (1982).
	\bibitem{mramond} C. Marshall, P.Ramond, Nucl.Phys. {\bf B85}, 375,
	(1975).
	\bibitem{hosocar} L.Carson, Y.Hosotani, Phys.Rev.D {\bf  37}, 1492
	(1988); ibid. 1519, (1988).
	\bibitem{ctuck} P.A. Collins, R.W. Tucker,
	Nucl.Phys. {\bf B112}, 150, (1976).
	\bibitem{minib} A. Aurilia, E. Spallucci,
	 Phys.Lett. {\bf B251}, 39, (1990);
	 \bibitem{minib2}
	 A. Aurilia, R.Balbinot,  E. Spallucci,
	  Phys.Lett. {\bf B262}, 222, (1991).
	\bibitem{dng} P.A.M.Dirac,
	Proc.\ Roy.\ Soc.\  {\bf A268}, 57 (1962).
	\bibitem{ht} P.S.Howe, R.W.Tucker,
    J.Phys.A {\bf 10}, L155, (1977).
\bibitem{netra} E.Eizenberg, Y. Ne'eman,
    Nuovo Cimento {\bf 102A}, 1183, (1989);\\
                B.P. Dolan, D.H. Tchrakian,
    Phys.Lett. {\bf B198}, 447, (1987).
\bibitem{gauge} A. Aurilia, A. Smailagic, E. Spallucci,
	Phys.Rev.D {\bf 47}, 2536, (1993).
	 \bibitem{gauge2}
	 A. Aurilia, E. Spallucci,
	  Class.Quant.Grav.{\bf 10}, 1217, (1993).
	 \bibitem{gauge3}
	  C. Castro, Int.J.Mod.Phys.{\bf A13}, 1263, (1998);\\
	J. Borlaf,
	Phys.Rev.D{\bf 60}, 046001, (1999);\\
	 D. B. Fairlie,
	 Phys.Lett.{\bf B484}, 333, (2000).
\bibitem{blt} J.A.deAzc\'arraga, J.M. Izquierdo, P.K. Townsend,
    Phys.Rev.D {\bf 45}, 3321, (1992);\\
              E.Bergshoeff, L.A.J. London, P.K. Townsend,
    Class.Quantum Grav. {\bf 9}, 2545, (1992).
\bibitem{eduard} E.I.Guendelman,
Class. Quantum Grav. {\bf 17}, 3673,  (2000).
	\bibitem{sec} S.Ansoldi, C.Castro, E.Spallucci,
	Class. Quantum Grav. {\bf 17}, 97,  (2000).
\bibitem{schild} A.Schild,
	Phys. Rev. D {\bf 16},  1722 (1977).
\bibitem{jacobi}
	C.Castro, ``{\it $p$-Brane Quantum Mechanical Wave Equations,}''
	hep-th/9812189;\\
	Y. Hosotani, R. Nakayama,
	Mod.Phys.Lett.{\bf A14}, 1983, (1999);\\
	L.M. Baker, D.B. Fairlie,
	Nucl.Phys. B{\bf 596},  348 (2001)
	\bibitem{eguchi} T.Eguchi,
	 Phys. Rev. Lett. {\bf 44}, n.3,  126 (1980).
	 \bibitem{sec2} S.Ansoldi, C.Castro, E.Spallucci,
	Class. Quantum Grav. {\bf 16}, 1833,  (1999).
	\bibitem{noi2}A.Aurilia,S.Ansoldi, E.Spallucci,
	Phys. Rev. D  {\bf 53}, n.2,  870 (1996).
	\bibitem{noi3}
	A.Aurilia, S.Ansoldi, E.Spallucci,
	Phys. Rev. D  {\bf 56}, n.4,  2352 (1997).
	\bibitem{noi4}
	A.Aurilia, S.Ansoldi, E.Spallucci,
	Chaos Sol. \& Fract. {\bf 10}, n.1, 1 (1998).
	 \bibitem{nrprop} A.Aurilia, S.Ansoldi, E.Spallucci,
	  Eur. J.Phys. {\bf 21}, 1 (2000).
	\bibitem{pezz}
	  W.M.Pezzaglia Jr. ``{\it Polydimensional Supersymmetric Principles}''
	   gr-qc/9909071;\\
	 W.M.Pezzaglia Jr. ``{\it Dimensionally Democratic Calculus and
	 Principles of Polydimensional Physics}''Proceed. of the {\it $5^{th}$
	 Int. Conf. on Clifford Algebras and their Applications in Math.
Phys.},
    Ixtapa--Zihuatanejo, Mexico, (1999), (R. Ablamowicz and B. Fauser eds.);
    gr-qc/9912025\\
	C. Castro, Jour. Chaos, Solitons and Fractals, Vol.11 (11) 1721
	(2000);\\
   C. Castro, ``The Status and Programs of the New Relativity Theory'',
   physics/0011040 v2, to appear in the Journ. Chaos, Solitons and Fractals.

	\end{thebibliography}
\end{document}